# Characterization of a Serializer ASIC chip for the upgrade of the ATLAS muon detector

Jinhong Wang, Liang Guan, Ziru Sang, J. W. Chapman, Tiesheng Dai, Bing Zhou, and Junjie Zhu

*Abstract*—We report on the design of a serializer ASIC to be used in the ATLAS forward muon detector for trigger data transmission. We discuss the performance of a prototype chip covering power dissipation, latency and stable operating line rate. Tests show that the serializer is capable of running at least at 5.76 Gbps with a bit error ratio below $1 \times 10^{-15}$, and a power consumption of 200 mW running at 4.8 Gbps. The latency between the start of loading 30 bits into the serializer to the transmission of the first bit from the serializer is measured to be about 6 ns.

## I. INTRODUCTION

THE Large Hadron Collider (LHC) [1] is the world's most powerful particle accelerator and provides unique opportunities for physicists to explore fundamental questions in particle physics. With the data taken in 2011 and 2012, both the ATLAS and CMS collaborations reported the discovery of a new particle that is consistent with the Higgs boson predicted by theorists as a key part of the Standard Model [2-3]. The LHC will be further increased to increase its instantaneous luminosity by a factor of 10 (up to $7 \times 10^{34}$ cm$^{-2}$s$^{-1}$) and will take data until about 2030, improving prospects for new discoveries. Experiments at the LHC will be upgraded periodically to deal with the increase in data rates and harsh radiation environment.

The ATLAS experiment [4] plans to replace the inner detectors of the end-cap muon spectrometer with the "new small wheel" detector (NSW) composed of eight layers of Micromegas chambers and eight layers of small-strip thin-gap chambers (sTGC) [5] in 2018. The sTGC is a gaseous wired proportional drift chamber and will be used for both trigger and precision tracking. Signals from the sTGC pad and strip detectors will be read out by the Amplifier Shaper Discriminator (ASD) ASIC [6], and then collected by the Trigger Data Serializer (TDS) ASIC [7] before being transmitted via twinax cables (up to 4 meters) to FPGA processing circuitry on the rim of the NSW detector. In order to reduce the output data rate, the TDS ASIC also performs pad-strip matching and only strips underneath certain pads will be read out.

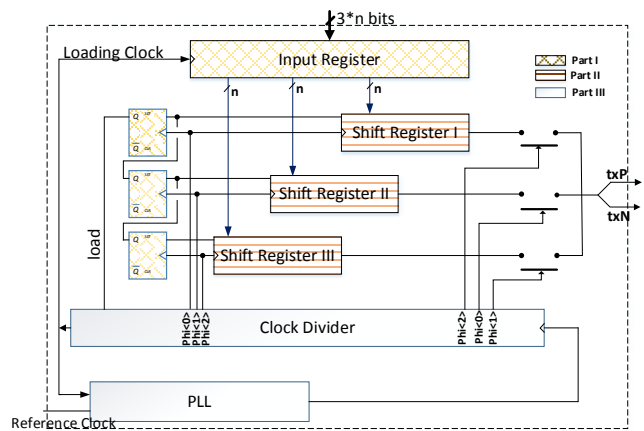

Fig. 1. Block diagram of the GBT serializer architecture, modified from [8].

The serializer section of the TDS ASIC is a crucial part of the TDS design and must meet requirements for speed (~5 Gbps), power consumption (< 0.5 W), radiation tolerance (300 kRad), and latency (< 10 ns and must be deterministic). We also want to take advantage of joint production with other ASICs in the same 130 nm CMOS technology within the NSW project (as [6]) to reduce the production cost. We did not find any commercial Intellectual Properties (IP) in the 130 nm CMOS process that meet all specifications, especially on the latency and radiation tolerance requirements. The Giga Bit Transceiver (GBT) serializer from CERN [8] uses the same 130 nm bulk CMOS process as [6] and is thus a good candidate. It features a 120:1 architecture and consumes about 0.3 W at 4.8 Gbps. The latency of the GBT serializer has not been reported but our simulation shows that the latency of only the serializer circuit is about 14 ns, neglecting any contribution from other processing logic. However, the GBT serializer does not satisfy all requirements needed for the serializer section of the TDS chip. It is also challenging to interface the serializer with the trigger processing logic circuits inside TDS for a low and deterministic latency as depicted in Section II.

In this paper, we present our modification to the GBT serializer to optimize performance for the ATLAS NSW sTGC trigger electronics. We evaluate the performance of a prototype of the modified serializer and discuss issues relevant to its applications.

The paper is organized as following: in Section II, we present the architecture and operation of the modified serializer.

Manuscript received on April 5th, 2015. This work is supported by the Department Of Energy under contracts DESC0008062 and DE-AC02-98CH10886.

The authors are with Department of Physics, University of Michigan, 450 Church Street, Ann Arbor, Michigan, 48109, US. (e-mail: jinhong@umich.edu; junjie@umich.edu).

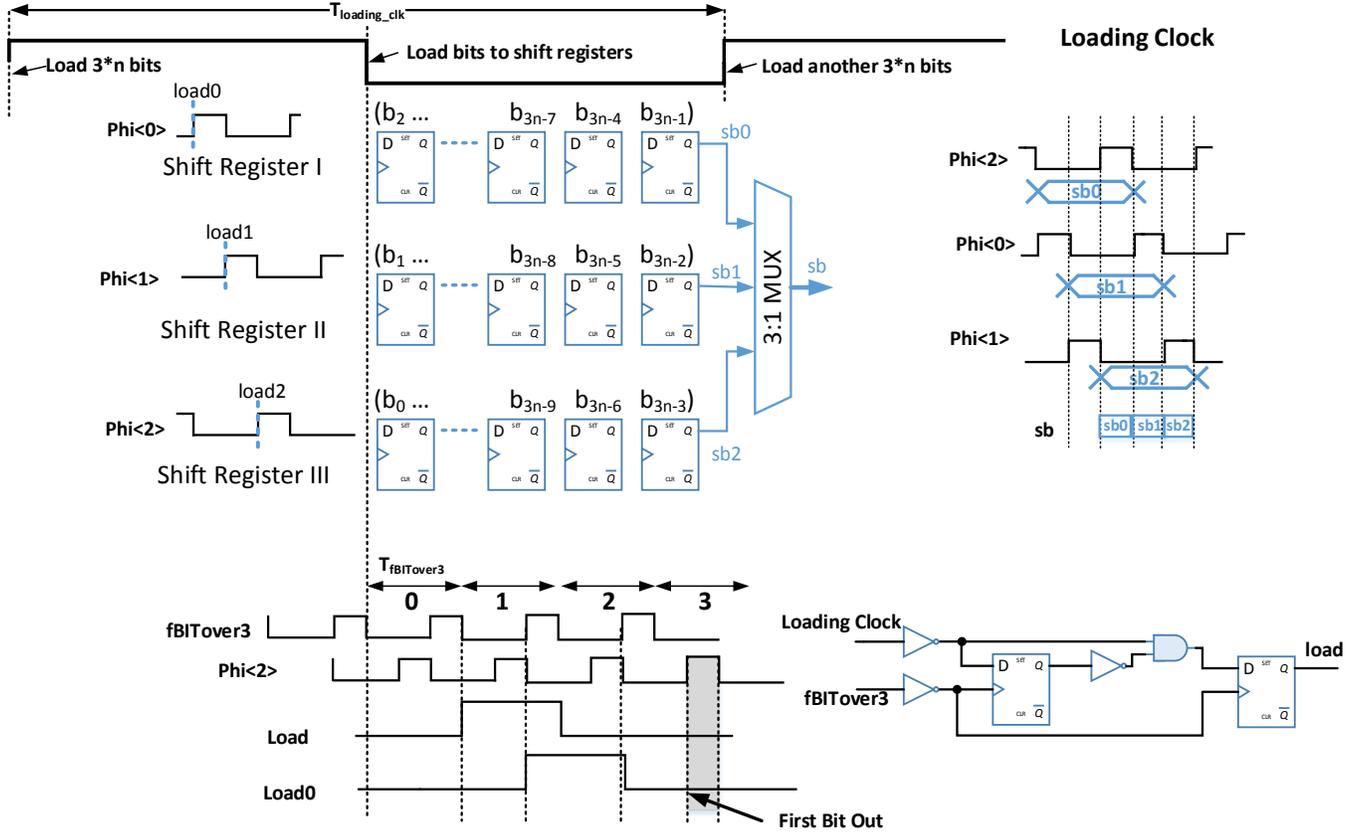

Fig. 2. Principle of operation for the GBT serializer. Diagrams have been simplified to highlight the operating principles.

In Section III, we characterize the serializer prototype. We discuss the performance in Section IV, and draw conclusions in Section V.

## II. The GBT Serializer In TDS

### A. Architecture and Principle of Operation

The GBT serializer is based on an interleaved architecture of three lines of shift registers, which work at one-third of the line frequency with 120-degree phase shift each other. The output of the shift registers are multiplexed to assemble the full line rate. A block diagram of the GBT serializer is shown in Fig. 1 [8] and can be divided into three parts. Part I is responsible for loading of the parallel bits. A total of $3 \times n$ bits are loaded into a register at the beginning of each loading clock cycle, as shown in Fig. 2. The $3 \times n$ bits, denoted as $b_i$ ($i$=0, 1, 2 …3n-1), are grouped by every three bits from Most-significant-Bit (MSB) to Least-significant-Bit (LSB) into three $n$-bits groups, which are then loaded into three shift registers in Part II. For example, the bits to Shift Register I are ($b_{3n-1}$, $b_{3n-4}$, $b_{3n-7}$ … $b_2$). These shift registers work at three different phases, Phi<$i$> ($i$=0, 1, 2), of a 1.6 GHz clock, with a phase difference of 120 degrees between each register. These Phi<$i$> clocks also exhibit a 1:2 high-low ratio, as shown in Fig. 2. In addition these clocks are used to control the loading of parallel bits into shift registers. Parallel bits to Shift Register I are loaded from the rising edge of Phi <0> after the falling edge of *Loading Clock* when the corresponding control signal *load0* is valid. Loading of parallel bits to Shift Register II and III is similar to the operation of Shift Register I, and is performed at the rising edge of Phi<1> with valid *load1* and the rising edge of Phi<2> with valid *load2*, respectively. Loading control signals of these registers, *load_k* (k=0, 1, 2), are registered copies of the *load* signal at the falling edges of Phi<2-$k$>. Generation of the *load* signal from the falling edge of *Loading Clock* is shown at the bottom of Fig. 2, where *fBITover3* is a copy of the 1.6 GHz clock with zero phase delay as Phi<0>. At the end of each line of the shift registers, a 3:1 multiplexer multiplexes the outputs, $sb_k$ ($k$ =0, 1, 2), to form the final 4.8 Gbps serial bit stream (*sb*). The multiplexing of the output of a shifter register line to *sb* is done at the previous phase of the 1.6 GHz clock with respect to its phase for the loading of parallel bits. For example, parallel bits to Shift Register I are loaded at Phi<0>, and *sb0* is multiplexed at the high point of the previous phase of Phi<0>: Phi<2>. Such configuration allows adequate setting time for $sb_k$ to be stable before multiplexing it to *sb*. Part III includes an on-chip PLL and generates all user clocks including the loading clock and Phi<$i$>. A detailed description of the design of this serializer can be found in [8].

### B. The serializer prototype in the TDS ASIC chip

The architecture of the GBT serializer can be summarized as loading $3 \times n$ bits at a speed of $f_{\text{load}}$ (*Loading Clock*), and serializing them out at a line rate ($f_{\text{line}}$) of $3 \times n \times f_{\text{load}}$. For the GBT serializer, $n$ =40, $f_{\text{line}}$ = 4.8 Gbps, and the loading speed is $f_{\text{load}}$ = 4.8 GHz /3/40 = 40 MHz which is the common clock frequency used at the LHC. For a constant line rate, a larger $n$ results in a lower $f_{load}$, and also a longer length of the minimum size of the data packet. For $n = 40$, the serializer loads 120 bits at a time and the minimum length of the data frame from the serializer is therefore 120. Inputs to the serializer will be frozen for a cycle of $f_{load}$ until the completion of the transmission of all 120 bits.

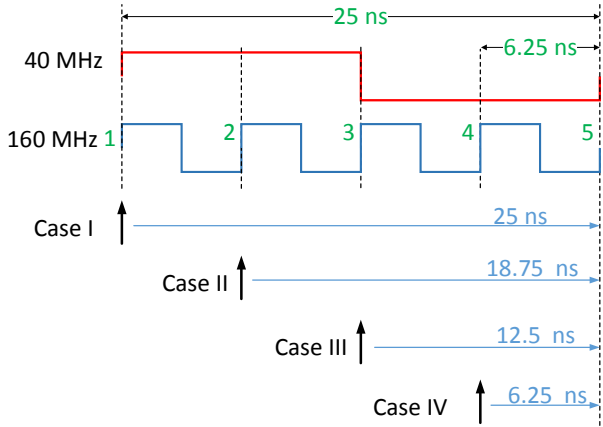

Fig. 3. Four phases possibilities of the 160 MHz clock to the 40 MHz clock inside TDS. Trigger data can be ready at any of the four cases.

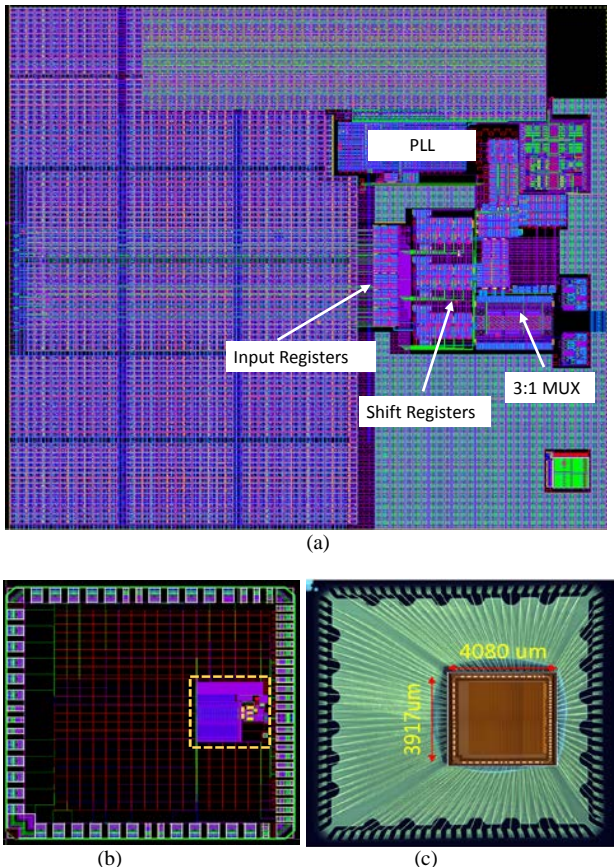

Fig. 4. Prototype of the modified serializer, where (a) shows the layout of the TDS serializer core, (b) is the layout of the serializer chip, (c) shows the wire-bonding diagram of the die to a 100-pin QFN package.

The modified serializer that we plan to use in the TDS ASIC follows the architecture in Fig. 1 but reducing the number of input bits from 120 to 30, thus reducing $n$ from 40 to 10. As a result, the serializer loads 30 bits at a rate of $f_{load} = 4.8$ GHz/3/10 = 160 MHz, rather than 120 bits at 40 MHz. There are two aspects motivating this modification. First, the trigger processing section preceding the serializer section in the TDS ASIC operates at 160 MHz. It is simpler to use the same clock to load the processed bits into the serializer to avoid issues of crossing between two frequency domains in one ASIC. Second, for the trigger application as the LHC experiments, low deterministic latency is of great concern. For the original GBT serializer, our simulation shows that it exhibits a latency of ~14 ns. Reducing the input bits from 120 to 30, lowers the latency to ~6 ns. Moreover, the TDS trigger processing section operates at 160 MHz and its latency may not be an integer multiplication of 25 ns, it is therefore possible for the trigger data to be ready at any of the four phases of a 40 MHz clock, as shown in Fig. 3. The 160 MHz and 40 MHz clocks both originate from the PLL of the serializer and their rising edges are aligned. In the worst scenario (case I), the trigger data (ready at clock tick #1) has to wait for 25 ns before being loaded into the serializer (at clock tick #5). This will make the total latency of the GBT serializer to be 14+25=39 ns. With the modified serializer presented here, the trigger data can be processed in the next 160 MHz cycle, and the corresponding latency is only ~6 ns (as described in the following section) plus one cycle of 160 MHz. The latency is defined as the interval between the time when the trigger data is ready in the trigger processing section and the time when the first bit of trigger data exits the serializer.

Modification of the serializer is mostly on Part I and Part II in Fig. 1. Length of the input register is trimmed to 30 bits, and as a result, the length of the shift register is reduced to 10 bits. The loading clock is now 160 MHz, four times that of the original counterpart. The loading control (load) signal is also adapted to load parallel bits into each register at 160 MHz. There are no modifications to the architecture of the PLL in Part III. Besides these architectural changes, we have also changed the metallization from a sub-process with 6 layers of thin metals, 2 layers of thick metals in the 130 nm CMOS process to a sub-process with 3 layers of thin metals, 2 layers of thick metals and 3 layers of thick metals for Radio Frequency (RF) applications. These changes are needed for joint production with other ASICs at ATLAS.

A prototype of the serializer has been fabricated to evaluate the performance of these modifications. Figure 4(a) shows the silicon layout of the modified serializer. The silicon area of the serializer core is about 1 mm × 1 mm, while the whole serializer die is 3.917 mm × 4.08 mm, as shown in Fig. 4(b). A 100-pin Quad Flat No-leeds (QFN) package was used for the prototype. The package size is 12 mm × 12 mm. Fig.4 (c) shows the wire-bonding diagram of the serializer die to the QFN package.

## III. CHARACTERIZATION OF THE SERIALIZER

A test board was built to evaluate the serializer prototype. Figure 5 shows a block diagram and a photograph of the serializer test board. The serializer inputs are provided by a Xilinx KC705 evaluation board [9] through a high-pin-count (HPC) connector. The reference clock of the serializer comes from an on-board PLL from Texas Instruments (*CDCE62005* [10]). The output frequency of the PLL is fully programmable for different test scenarios. There is also a copy of the PLL output for test purposes. The output of the serializer can be sent to test equipment like an oscilloscope with a pair of SMA connectors.

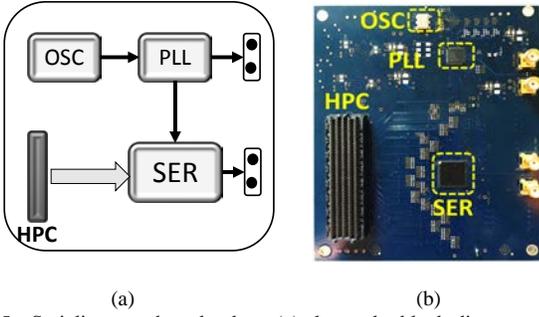

Fig. 5. Serializer test board, where (a) shows the block diagram of the test board, and (b) is a picture of the test board.

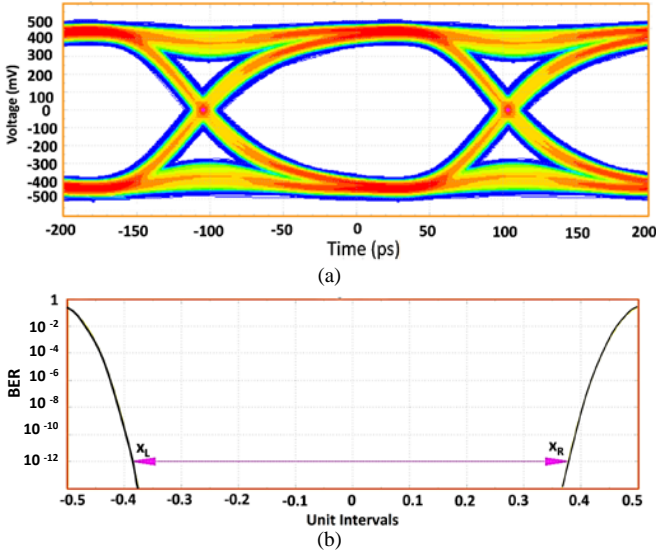

Fig. 6. Performance of the modified serializer at 4.8 Gbps, where (a) shows an eye diagram; (b) is a bathtub curve evaluated from the oscilloscope.

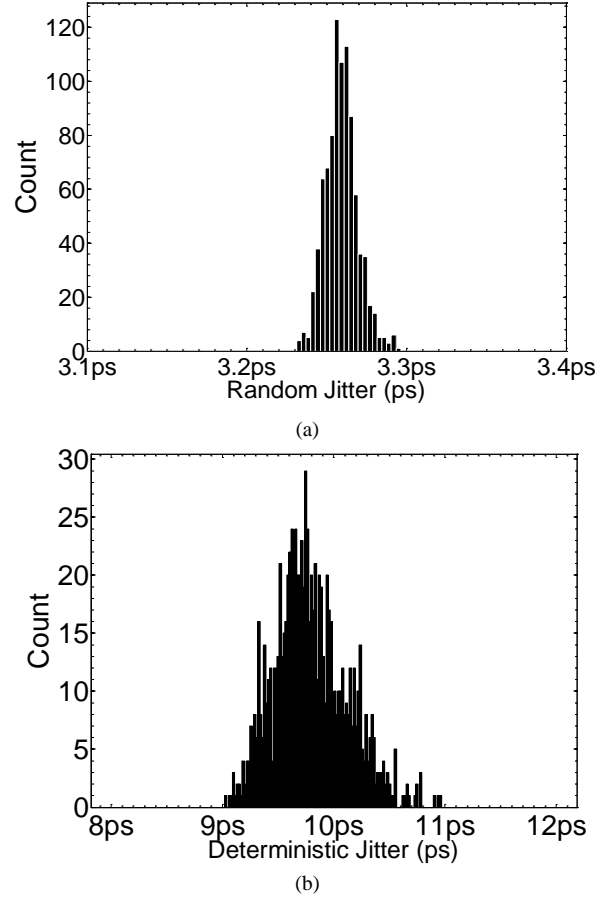

Fig. 7. Statistical analysis of jitter at 4.8 Gbps, where (a) shows distribution of random jitter; (b) is histogram of deterministic jitter.

### A. Performance of the serializer

An eye diagram is a good method to represent and analyze a high-speed signal. It allows a quick visualization and determination of key parameters of the serializer. In our test, the eye diagram is evaluated with a Tektronix 12.5 GHz bandwidth 50 GS/s oscilloscope (*DSA71254B* [11]). The input to the serializer is a pseudorandom binary sequence with every permutation of 31 bits (PRBS-31) generated from a Xilinx KC705 FPGA evaluation board. The PRBS-31 is based on a linear feedback shift register with a polynomial $x^{31} + x^{28} + 1$.

Figure 6(a) shows the eye diagram of the serializer output at 4.8 Gbps over a one-meter 50 $\Omega$ coaxial cable. The height of the eye at the center is about 720 mV and the shape indicates good signal integrity from the serial output. A bathtub diagram of the bit-error ratio (BER) is evaluated from the eye samples in the oscilloscope, as shown in Fig. 6(b). The $x$ coordinate is number of Unit Intervals (UI) with one UI defined as width of a bit at a line rate. The width of the eye is calculated as that of the bathtub curve at a BER of $1 \times 10^{-12}$, which is 0.76 UI at 4.8

Gbps, corresponding to about 158.6 ps. The Total Jitter (TJ$_{BER}$) is estimated using (1):

$$TJ_{BER} = UI - (x_R - x_L) \tag{1}$$

where $x_R$, $x_L$ are the right and left horizontal coordinates of the bathtub curve with a BER equals to $1 \times 10^{-12}$ in Fig. 6(b), respectively.

From (1), the total jitter is estimated as 0.24 UI, or about 49.7 ps. Total jitter includes unbounded and bounded components. The unbounded component is also called Random Jitter (RJ) and is unpredictable. It is usually due to thermal noise or other uncorrelated noise sources, and can be measured as the RMS of all timing errors exhibiting deterministic behavior. The bounded component is deterministic and is termed Deterministic Jitter (DJ). DJ usually has a specific cause such as cross-talk and simultaneous outputs switching, and is quantized as the peak-to-peak amplitude for all timing errors that follow deterministic behavior. DJ can be further sub-classified into Period Jitter (PJ) and Data-Dependent Jitter (DDJ), in which PJ repeats in a cyclic fashion whereas DDJ is bit-sequence correlated. A complete jitter analysis using the jitter analysis package (*DPOJET* [12]) embedded in the oscilloscope shows that there is about 3.26 ps random jitter (RJ) and about 9.78 ps deterministic jitter (DJ). Corresponding distributions of RJ and DJ are shown in Fig. 7(a) and (b), respectively. The RMS of RJ is about 10 fs, whereas that of DJ is about 320 fs. We also observe that DDJ is almost zero, and PJ occupies nearly 100% of the DJ. PJ is typically caused by external deterministic noise sources coupled to the test setup, such as the switching noise on the power supply. It could also come from the clock recovery scheme of PLL in the oscilloscope.

We utilize the embedded PRBS checker inside the Xilinx 7 Series FPGA [13] to evaluate BER of the serializer. The test is performed using the same PRBS-31 pattern and the setup achieves an error-free running of more than 3 days, corresponding to a BER that is less than $1 \times 10^{-15}$.

### B. Latency of the Serializer

The serializer loads 30 bits at the rising edge of *Loading Clock*, as shown in Fig. 2. These 30 bits are then divided into three 10-bits groups and loaded into three shift registers after the falling edge of *Loading Clock*. For *Shift Register I*, the 10 bits are loaded at the rising edge of Phi<0> when *load0* is valid, and the first bit is out of the 3:1 multiplexer at high of the following Phi<2>. The total time ($t_{delay}$) from the rising edge of *Loading Clock* to the time that the first bit is out is calculated as

$$t_{delay} = T_{loading\_clk}/2 + (3+2/3) \times T_{fBITover3}$$
$$= (3.125 + 2.292) \text{ ns} \approx 5.42 \text{ ns} \qquad (2)$$

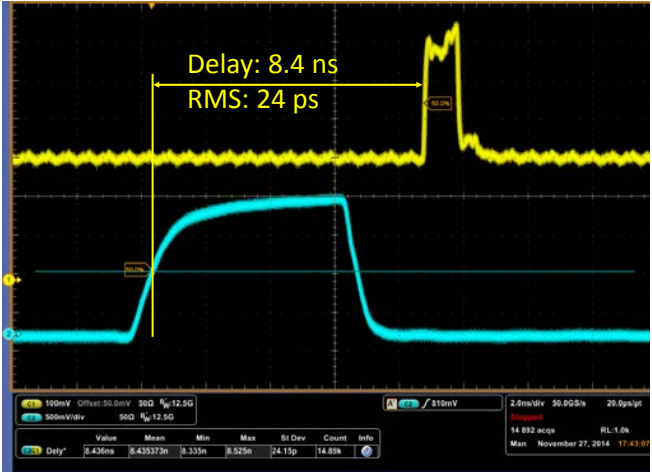

Fig. 8. Measurement of the serializer latency on a DSA71254B oscilloscope.

where $T_{loading\_clk}$ and $T_{fBITover3}$ are periods of *Loading Clock* and *fBITover3*, respectively. The expression in (2) represents the internal latency of the serializer.

In our test, we evaluate the latency by using a KC705 evaluation board to feed the serializer with a 150-bit pattern starting with 5 consecutive one bits and followed by 145 zero bits. The pattern is sent in five consecutive 30-bit frames with the 5 one bits coming out first as the most significant bits. In this way, we will see a periodic pulse of about 1 ns width at the serializer output every five cycles of *Loading Clock*, i.e., every 31.25 ns. A pulse of one *Loading Clock* width is also generated as a probe from the evaluation board every time the pattern is produced, and this pulse signal is sent to an on-board SMA connector. We use coaxial cables with equal length to send the probe signal and the serializer output to the oscilloscope, and calculate the latency as the time difference between the rising edges of the probe signal and the serializer output signal.

Figure 8 shows the oscilloscope measurement where the top is the serial output, and the bottom is the probe pulse. The test is done by periodically resetting the serializer every 3 seconds and the oscilloscope persists all pulses on the screen for monitoring. We used the embedded statistical analysis function in the oscilloscope to analyze the measured delay ($t_{measure}$), and observed that the mean is 8.4 ns with a variation of 24 ps RMS. There are no large variations observed on the delay from the persistence of waveforms or from the recorded minimum and maximum values. We therefore conclude the serializer features a fixed latency.

The measured latency is different from that shown in (2), and the difference could be interpreted by breaking down the measured latency. The time interval from the generation of the probe pulse to the time being captured on the oscilloscope ($t_{probe}$) includes the propagation delays of the FPGA IOs ($t_{IO}(probe)$), PCB traces on the evaluation board ($t_{PCB}(probe)$), and the coaxial cable ($t_{cable}(probe)$), as shown in (3):

$$t_{probe} = t_{IO}(probe) + t_{PCB}(probe) + t_{cable}(probe) \qquad (3)$$

The time interval from loading of the first 30 bits of a 150 bits pattern (from MSB to LSB) to the time the serial output pulse being captured by the oscilloscope ($t_{SER}$) includes the phase difference of the 160 MHz clock in the FPGA to the phase of 160 MHz inside the serializer ($t_{phase}$), the propagation delay of FPGA IOs and serializer IOs ($t_{IO}(SER)$), the propagation delay of traces on the evaluation board, the serializer test board, and the HPC connector ($t_{PCB}(SER)$), and that of the coaxial cable ($t_{cable}(SER)$), as shown in (4):

$$t_{SER} = t_{phase} + t_{IO}(SER) + t_{PCB}(SER) + t_{cable}(SER) + t_{delay} \qquad (4)$$

Therefore we have:

$$t_{measure} = t_{SER} - t_{probe}$$
$$= t_{delay} + t_{phase} + (t_{IO}(SER) - t_{IO}(probe))$$
$$+ (t_{PCB}(SER) - t_{PCB}(probe)) \qquad (5)$$

In (5), we assume $t_{cable}(probe) = t_{cable}(SER)$, and $t_{phase}$ is in the range of (0, 6.25 ns). Therefore it is possible that the measured latency is larger than that estimated in (2) as additional delays are introduced due to phase difference and propagation delay difference.

### C. Power Consumption

We measured the power consumption of the serializer by cascading a 0.15 Ω resistor in serial to the 1.5 volt supply ($V_{SER}$) of the serializer. The total power consumption of the serializer is the sum of static power and dynamic power consumption. Static power is approximated while holding the serializer in reset. The corresponding current consumption for the nominal operation and the reset is 116 mA and 55.3 mA respectively. The power is therefore 174 mW and 83 mW respectively. It is important to keep the serial resistor as small as possible so that the voltage drop across it is negligible.

## IV. DISCUSSION

### A. Radiation Tolerance

Radiation tolerance is another concern for the serializer to be integrated in the NSW trigger electronics, as the TDS ASIC sits on the front-end board on the detectors. The Total Ionizing Dose (TID) for the NSW sTGC trigger front-end boards is about 300 kRad with a safety factor of 6 [14]. This level of radiation is not an issue for the 130 nm CMOS process we used [15]. However to protect the TDS ASIC (including the serializer) from unstable operations or permanent failures requiring recovery from a reset or reconfiguration in case of Single Event Upsets (SEU), we adapt all the Triple Modular Redundancy (TMR) protection schemes of the PLL from the original GBT serializer. There is a single event transient simulation in [8].

### B. Serializer line rate

The serializer is targeted to work at 4.8 Gbps, however, there is flexibility to run it at a higher data rate. We changed the reference clock to the serializer from 40 MHz to 48 MHz and 53.33 MHz, which corresponds to an output data rate of 5.76 Gbps and 6.4 Gbps, respectively. We repeated the evaluation procedure for these two line rates with the oscilloscope and the embedded PRBS checker in a Xilinx 7 series FPGA. We found that there are no problems for the serializer to work at 5.76 Gbps. However, the 6.4 Gbps output is not stable and the link failed with a high BER. We therefore estimate the highest achievable rate is between 5.76 Gbps and 6.4 Gbps.

The modified serializer can also be used in situations where a lower line rate is expected. For these applications, one can use a lower reference clock but will be limited by the minimum working frequency of the serializer. Simulation of all condition corners indicates that the lowest line rate is between 0.5 and 3.5 Gbps. We have tested four evaluation boards and our results indicate that the serializer is still stable at 1.44 Gbps, which is the lowest line rate that we can evaluate using our evaluation board. One can also send the same bit multiple times to reduce the effective line rate. For example, supposing $\{b_i\}$ ($i = 0,1,2…$) are the bits to be loaded into the serializer at 4.8 Gbps. If we double the bits, i.e. $\{b_i\ b_i\}$, the effective output rate will be halved: 2.4 Gbps.

### C. Latency of the Serializer in TDS

The latency of the serializer core is estimated to be around 6 ns, whereas that of the original GBT serializer core is about 14 ns. If we integrate the GBT serializer core into TDS, the total latency could be as large as 39 ns when the trigger data crossing from the 160 MHz clock domain to the 40 MHz clock domain. This large latency is unacceptable for the trigger application. We compare the latency and other parameters of the two serializers in Table I. For the modified serializer, there is only one clock domain and the trigger data is loaded right in the next 160 MHz cycle. The expected latency is ~6+6.25=12.25 ns. Whereas for the original GBT serializer, the latency depends on the time when the trigger data is ready: in the best case (Case IV in Fig.3) the latency is around 14+6.25=20.25 ns; while in the worst scenario (Case I in Fig. 3), the latency can be ~14+25=39 ns.

Besides, our test also shows that the serial core latency is deterministic. Deterministic latency is preferable in serial transmission applications such as the ATLAS sTGC trigger system, where the receiving end requires a predictable arrival time of information. The scheme of deterministic latency of the serializer comes from the fact that the serializer always loads data with respect to the rising edge of the 160 MHz *Loading Clock*, and there is a deterministic phase relation between *Loading Clock* and the reference clock.

## V. CONCLUSION

A fast-speed low-latency radiation-tolerant serializer is needed for the upgrade of the ATLAS NSW trigger electronics. We modified the CERN GBT serializer to load 30 bits at 160 MHz and changed its metallization for joint production with other NSW ASIC chips. A serializer prototype was fabricated and its performance was evaluated. Our tests show that the serializer is capable of running at 4.8 Gbps with a BER less than $1 \times 10^{-15}$. The highest achievable rate is between 5.76 Gbps and 6.4 Gbps. The power consumption of the serializer in nominal operating conditions (1.5 volt supply at about 25 °C ambient temperature) is about 200 mW. The latency is measured to be less than 10 ns, and is fixed upon reset or power up operations. The prototype chip has the potential to be used as a separate chip or as an IP in applications that need high speed data transmission with low and deterministic latency in a high radiation environment.

### ACKNOWLEDGMENTS

The authors would like to thank Paulo Moreira, Rui Francisco, Filip Tavernier from CERN, De Geronimo, Gianluigi from Brookhaven National Laboratory, T. Liu and D. Gong from Southern Methodist University, Edward Diehl from the University of Michigan for their help in this work; and the reviewers, Editor and Senior Editor for valuable suggestions they provided during the review process.

TABLE I
COMPARISON BETWEEN THE MODIFIED SERIALIZER AND THE ORIGINAL GBT SERIALIZER.

| Catalog | Modified Serializer | GBT Serializer |
|---|---|---|
| Technology | 130 nm bulk CMOS | 130 nm bulk CMOS |
| Speed | 4.8 Gbps | 4.8 Gbps |
| Architecture | 30 : 1 | 120 : 1 |
| User clock | 160 MHz | 40 MHz |
| Power | ~ 0.2 Walt | ~0.3 Walt |
| Radiation Tolerance | Yes | Yes |
| Latency of Serializer Core | ~6 ns | ~14 ns |
| Latency when integrated in TDS[a] | ~12.25 ns | ~20.25 ns (min.) ~39 ns (max.) |

[a] The latency is defined as the time interval between the time when the trigger data is ready and the time when the first bit exits the serializer